\documentclass[10pt,prl,twocolumn,showpacs]{revtex4-1}
\usepackage{amsmath}
\usepackage{graphicx}
\usepackage{amsfonts}
\usepackage{amssymb}
\usepackage{color}
\usepackage{enumitem}
\usepackage{multirow}

\begin{document}

\title{
Kramers' doublet ground state in topological Kondo insulators
}
\author{M.~A.~Griffith$^1$, M. A. Continentino$^1$, and  T.~O.~Puel$^{2,3,1}$ }
\email{tharnier@csrc.ac.cn}

\affiliation{
$^1$Centro Brasileiro de Pesquisas F\'isicas, Rua Xavier Sigaud 150, 22290-180, Rio de Janeiro, Brazil\\
$^2$Beijing Computational Science Research Center, Haidian District, Beijing 100094, China\\
$^3$CeFEMA, Instituto Superior T\'ecnico, Universidade de Lisboa, Av. Rovisco Pais, 1049-001 Lisboa, Portugal
}

\date{\today}


\begin{abstract}
We consider the simplest variant of a Kondo insulator where a doublet of
localized $f$-electrons hybridizes with spin-degenerate conduction electrons.
We analyse the symmetries of $f$-orbitals involved in the hybridization
and point out that the effective four-band model
of such systems provides further descriptions of clean Kondo insulators,
namely the spin-texture of the surface states are described by a $\mathbb{Z}$ topological invariant.
We discuss general conditions for the appearance of topological non-trivial states and implications for rare-earth based compounds.
As an example, we derive the full phase diagram of tetragonal Kondo insulators.
In particular, our findings describe the spin-textures in the physically interesting non-trivial topological phase,
i.e. when the band-width of conduction electrons sets the largest energy scale,
and a new weak topological phase appears as function of the normalized distance between bands' centers.
\end{abstract}

\maketitle

\section{Introduction}
Kondo insulators recently have attracted a lot of attention due to their promise to realize topological phases with a large bulk gap generated by strong electron correlations~\cite{Dzero2010,Takimoto2011,Dzero2012,Carpentier,Dzero2013,Kim2014,XiangLi2014,DzeroReview2015}.
Different effective models have been proposed for several candidate materials, but not all of them are in a strong topological phase protected by a non-trivial
$\mathbb{Z}_2$-invariant~\cite{Alexandrov2013,PhysRevB.89.085110,PhysRevB.90.201106}.
In the context of ${\rm SmB}_6$, one promising candidate for a topological Kondo insulator, the consequences of mirror symmetries have been pointed out~\cite{Vojta2015,PhysRevLett.115.156405}.
The latter allow for a refined topological characterization and reflects in the surface states spin-structure,
for instance, spin expectation value of surfaces were observable in ${\rm SmB}_6$ from spin-resolved ARPES experiments~\cite{Xu:2014aa}.
Improving the topological characterization of Kondo insulators is, from a broader perspective, relevant for the identification of further promising materials.

Here, we revisit the simplest variant of a three-dimensional
Kondo insulator where
a doublet of localized $f$-electrons hybridizes
with spin-degenerate conduction electrons.
We point out that not only the lattice-symmetry of the material,
but also
the symmetry of the $f$-orbitals involved in the hybridization
can allow
for an improved topological
characterization of the Kondo insulator, 
which results from a rotational invariance of
the involved orbital wave-functions.
Specifically, we show that the topological properties of insulators
involving localized Kramers' doublets of lowest
angular momentum projection,
${\Gamma^J_{1/2} = |J,m_J=\pm1/2\rangle}$,
can be understood from a fine-tuned Hamiltonian characterized
by a $\mathbb{Z}$-invariant.
This work enlighten this connection and the conditions to relate the $\mathbb{Z}$ invariant
to the spin-textures of the $\mathbb{Z}_2$ Kondo insulators.

For tetragonal Kondo insulators we show that it is particularly useful
when the band-width of conduction electrons sets the largest energy scale.
On the other hand, when other than the $\Gamma^J_{1/2}$-doublet participate in the hybridization,
it may only appears in a low hopping neighbor expansion,
i.e., it is broken by higher order neighbor contributions 
(which depend on both the involved doublet and crystal symmetry).
This property, therefore, can only exist in those crystalline lattice structures
that allow for a pure $\Gamma^J_{1/2}$-doublet in the ground state.

We also identify the relevant
point-group symmetries for Kondo insulators
involving
doublets from the $J=5/2$- and 7/2-multiplets
and
discuss 
implications for rare-earth based compounds.

The sections are organized as following:
next section we introduce the model,
in section \ref{Kramers doublets} the special symmetry of $f$-orbitals is discussed,
as well as the equivalent symmetry in low order neighbor approximation.
In section \ref{connected hamiltonian} we detail the connection between Hamiltonians in different classes,
and how their topological invariant are related, i.e. the improved characterization of topological Kondo insulators.
Sections \ref{applications} and \ref{discussion} discuss the implications for rare-earth based compounds
and resume the finds in this work, respectively.

\section{Model}
\begin{figure}[t]
  \centering
  \includegraphics[width=0.9\linewidth]{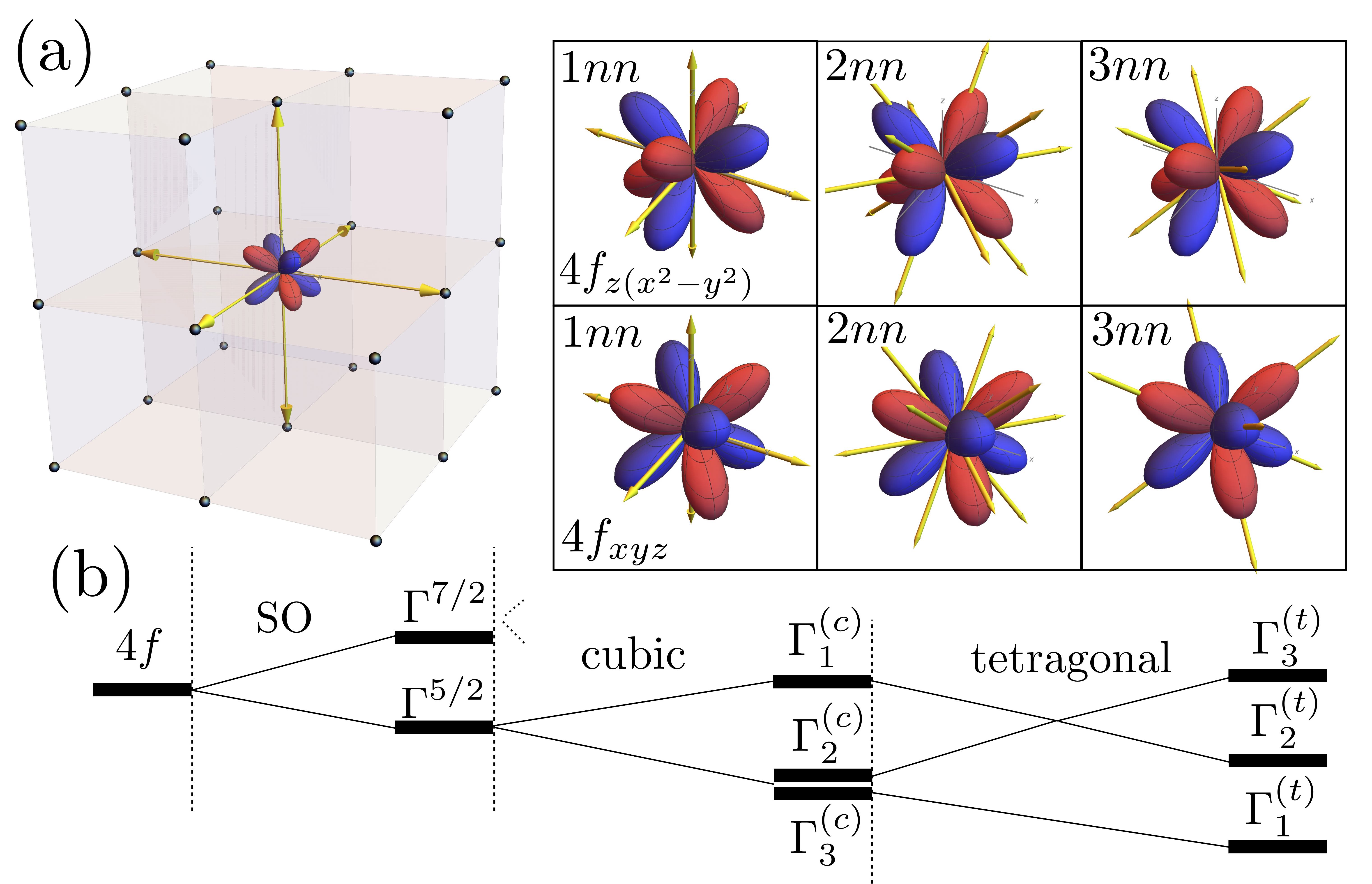}
  \caption{(Color online)  
  The low-order neighbor expansion is illustrated in panel (a), where the arrows point to the next-neighbor directions in the cubic lattice, which is filled with the orbital configurations $4f_{xyz}$ and $4f_{z(x^2 - y^2)}$.
  For the first next-neigbors ($1nn$) one notice that their directions don't coincide with the atomic orbitals,
  while for $2nn$ and $3nn$ they coincide with the $4f_{z(x^2 - y^2)}$ and $4f_{xyz}$, respectively.
  Panel (b) shows the $f$ orbital degeneracy splitting caused by strong spin-orbit (SO) coupling followed by crystalline field.}
  \label{fig2Orbital}
\end{figure}

We start out from the simplest variant of a
$3D$ Kondo insulator,
where a spin-degenerate wide conduction band
hybridizes with a narrow band formed by degenerate doublets $\Gamma^J = |J, \pm \rangle$ of nearly localized $f$-electrons,
\begin{align}
\label{Hamiltonian}
\hat H
& =
\sum_{\bold{k}}\left(
\sum_{\sigma=\uparrow,\downarrow}
\varepsilon_{\bold{k}}^{c}c_{\bold{k},\sigma}^{\dagger}c_{\bold{k},\sigma}
+
\sum_{s=\pm} \varepsilon_\bold{k}^{f} f_{\bold{k},s}^{\dagger}f_{\bold{k},s}\right)
\nonumber\\
&\,\,\,
+
\sum_{\bold{k}}\sum_{\sigma=\uparrow,\downarrow}\sum_{s=\pm}
\left(V_{\bold{k}, \sigma s}c_{\bold{k},\sigma}^{\dagger}f_{\bold{k},s}
+
V_{\bold{k}, \sigma s}^{*}f_{\bold{k},s}^{\dagger}c_{\bold{k},\sigma}\right).
\end{align}
Here
$\varepsilon_\bold{k}^{c,f}$ are the energy-dispersions
of conduction and $f$-electrons, respectively, and
$V_{\bold{k}, \sigma s}$ account for their hybridization
(both, $\varepsilon_\bold{k}^{f}$ and $V_{\bold{k}, \sigma s}$ are considered as effective parameters that include effects from electron correlations.
Treating correlations beyond the mean-field limit is  very challenging, particularly in three dimensions. Most of these approaches consider one or two-dimensional systems
\cite{PhysRevB.88.235111,PhysRevB.87.165109,PhysRevB.88.035113,2015arXiv150703477B,PhysRevB.93.125103,PhysRevX.5.021017,refId0}).
Throughout this work we always
assume a sufficiently large crystal field, which
separates a Kramers' degenerate ground-state 
from the $5/2$- or $7/2$-multiplet. 
While generally $\Gamma^J$
is 
some linear combination of the angular momentum eigenstates
$\Gamma^J_{m_J}\equiv |J,  \pm m_J\rangle$,
of specific interest to us are
cases in which $\Gamma^J =\Gamma^J_{1/2}$.
Fig.~\ref{fig2Orbital}b illustrates the crystal field splitting after the addition of spin-orbit for the cubic and tetragonal structures,
where $\Gamma^{(c),(t)}_{1,2,3}$ follow the notation in Ref.~[\onlinecite{Dzero2012}],
indeed the interesting ground state here is $\Gamma^{(t)}_{1}=\Gamma^{5/2}_{1/2}$.
Before we go into details,
it is convenient to express Eq.~\eqref{Hamiltonian} in the matrix form
$\hat H=\sum_{\bold{k}}
\Psi_{\bold{k}}^\dagger
{\cal H}(\bold{\bold{k}})
\Psi_{\bold{k}}$, where
${ \Psi^\dagger_\bold{k}
=(c_{\bold{k}\uparrow}^\dagger, c_{\bold{k}\downarrow}^\dagger,
f_{\bold{k},+}^\dagger, f_{\bold{k},-}^\dagger) }$
and
\begin{equation}
\label{HamiltonianDirac}
  {\cal H}(\bold{k})
  = \sum_{i=0}^5 h_i(\bold{k})\gamma_i,
\end{equation}
with
$h_{0,4}(\bold{k})=(\varepsilon_{\bold{k}}^{c}\pm\varepsilon_\bold{k}^{f})/2$
and remaining coefficient functions
$h_{i}(\bold{k})$ defined by the hybridization elements $V_{\bold{k},\sigma s}$.
Here $\gamma_0=\openone_4$ is the identity matrix and
$\gamma_{i}=\sigma_{i}\otimes\tau_1$ (for $i=1,2,3$),
$\gamma_4= \openone_2\otimes\tau_3$,
and
$\gamma_5= \openone_2\otimes\tau_2$,
are Dirac matrices
satisfying the Clifford algebra
$\{ \gamma_a,\gamma_b\} = 2\delta_{ab}$, with
Pauli-matrices $\sigma_i$ and $\tau_i$ operating in spin- and orbital-space, respectively.
The general form of Eq.~\eqref{HamiltonianDirac} is fixed by invariance under
inversion, ${\cal I}=\sigma_0\otimes\tau_{3}$,
and time-reversal, ${\cal T}= i\sigma_{2}\otimes\tau_0\,K$
(with $K$ the complex conjugation)~\footnote{
Inversion symmetry
 ${\cal I}{\cal H}_{\bold k}{\cal I}^{-1}={\cal H}_{-\bold k}$
restricts the diagonal blocks of ${\cal H}_{\bold k}$ in orbital-space to be even,
and off-diagonal hybridization blocks to be odd functions in $\bold{k}$.
In presence of time reversal-symmetry,
${\cal T}{\cal H}_{\bold k}{\cal T}^{-1}={\cal H}_{-\bold k}$,
this excludes Pauli-spin matrices on the orbital diagonal blocks of ${\cal H}$, while
restrictions on hybridization blocks
are different for
symmetric and anti-symmetric hybridization combinations
$c^\dagger f \pm f^\dagger c$, i.e.
symmetric elements involve Pauli spin-matrices while anti-symmetric elements
are independent of spin orientations.
Notice that the choice of a relative phase-factor between $c$ and $f$-electrons,
for which ${\cal T}$ is of the indicated form,
is reflected in coefficients $c_m,{\bar c}_m$ in Eq. (\ref{vhy}), 
and one may alternatively work in a basis where
 ${\cal T}=\tau_3\otimes i\sigma_{2}K$.}.
In some cases, the symmetry of participating $f$-orbitals 
imposes an additional constraint to Eq.~\eqref{HamiltonianDirac} as we are going to discuss next.

\section{Kramers' doublets $\Gamma^{J}_{1/2}$}
\label{Kramers doublets}
To illustrate the point consider the hybridization block
$c^\dagger_\bold{k} V_{m_J} (\bold{k}) f_\bold{k}$
for one of the
Kramers' doublets
$\Gamma^{5/2}_{m_J}$.
Following previous work~[\onlinecite{Yamada2012}]
the $2\times2$ hybridization matrix
reads 
\begin{align}
\label{vhy}
V_{n-\tfrac{1}{2}}(\bold{k})&=
\begin{pmatrix}
c_n\, {\cal Y}^{n-1}_3(\bold{k})
&
\bar c_n\, {\cal Y}^{-n}_3(\bold{k})
\\
- \bar c_n\, {\cal Y}^{n}_3(\bold{k})
&
- c_n\, {\cal Y}^{-n+1}_3(\bold{k})
\end{pmatrix},
\end{align}
where $c_n$ and $\bar c_n$ are purely real/imaginary numbers
for $n$ even/odd (as fixed by time-reversal symmetry) 
and
\begin{align}
\label{calY}
{\cal Y}^m_3(\bold{k})
=
\sum_{\bold{R}\neq 0} v(|\bold{R}|)\,
Y^m_3(\hat{\bold{R}})
e^{i\bold{k}\cdot \bold{R}}.
\end{align}
The sum in \eqref{calY} runs over all neighbor sites $\bold{R}$,
$Y_3^m(\hat{\bold{R}})$ are the spherical harmonic functions of
$f$-orbitals (with $\hat{\bold{R}}$ a unit vector),
and coefficients  $v(|\bold{R}|)$ depend on the neighbor-distance.
A similar expression~\eqref{vhy} holds for the Kramers' doublets
$\Gamma^{7/2}_{m_J}$ $\,$ \footnote{I.e. without negative signs in the first line, and
$c_n$, $\bar c_n$ now purely real/imaginary numbers for $n$ odd/even} and the following
discussion therefore applies to both multiplets $J=5/2$ and $J=7/2$.

Recalling that ${\cal Y}_3^{-m}(\hat{\bold k})=(-1)^{m+1} [{\cal Y}_3^{m}(\hat{\bold k})]^*$,
it is verified
that Eq.~\eqref{vhy} involves four independent real-valued functions.
This is tantamount to noting that in general 
the hybridization block~\eqref{vhy}
requires in the Hamiltonian~\eqref{HamiltonianDirac}
 a linear combination 
of the four matrices $\gamma_{1,2,3,5}$.
A different situation, however, occurs for the Kramers' doublet $\Gamma^J_{1/2}$
where the hybridization block involves the spherical harmonic
$Y_3^0(\hat{\bold{R}})$ on its diagonal.
Rotational symmetry of the latter implies that $c_n{\cal Y}_3^0(\bold{k})$ is a purely real-valued function and
Eqs.~\eqref{vhy} and~\eqref{HamiltonianDirac} are spanned by only three out of the four independent $\gamma$-matrices,
i.e., $\gamma_{1,2,3}$~\footnote{For $J=7/2$ it is the matrices $\gamma_i$ with $i=1,2,5$.}.
As we discuss in the next section,
the remaining matrix $\gamma_5$ is crucial to the improved characterization of the effective Hamiltonian~\eqref{HamiltonianDirac}.

\subsection{Low neighbor expansion}
First, we notice that
Eqs.~\eqref{HamiltonianDirac} and~\eqref{vhy} with only three of the four $\gamma$-matrices
may also appear in a low-order neighbor expansion
for other than $m_J = 1/2$, but in this case it is not a robust constraint.
For illustration consider Eqs.~\eqref{vhy} and \eqref{calY}
 in a cubic environment for the
 Kramers' doublets $\Gamma^{5/2}_{3/2}$,
\begin{align}
V_{\tfrac{3}{2}}(\bold{k})
& \propto
\begin{pmatrix} h_3(\bold{k})-ih_5(\bold{k}) & h_1(\bold{k})-ih_2(\bold{k})\\
h_1(\bold{k})+ih_2(\bold{k}) & -h_3(\bold{k})-ih_5(\bold{k})
\end{pmatrix}.
\label{genericHybridization}
\end{align}
While coefficient functions $h_3$, $h_5$ are non-vanishing already for nearest neighbors,
 $h_1$, $h_2$ become finite only starting from second and third order neighbors, respectively.
  This vanishing of $h_1$, $h_2$
  is here traced back to the specific values of spherical harmonics $Y_3^{\pm 2}$ 
  at the angles of the near neighbor-directions in the cubic lattice, as illustrated in  Fig.~\ref{fig2Orbital}a
   (i.e. zero for nearest and purely real for next-nearest neighbors),  
 and also holds for tetragonal or orthorhombic but e.g. not hexagonal lattices.
The absence of $h_5$ in case of the $\Gamma^J_{1/2}$ doublet discussed above,  
on the other hand, follows from
the rotational symmetry of involved orbital functions and, therefore, applies for 
all neighbor contributions.

\section{Fine-tuned Hamiltonian}
\label{connected hamiltonian}
We first note that in the translational invariant insulating phase, i.e. clean system, one can always remove $h_0 (\bold{k})$ from Eq.~\eqref{HamiltonianDirac}
without closing the gap.
In addition, the topological phase diagram remains unaltered
since eigenfunctions are not affected by terms proportional to identity.
In general this procedure describes an adiabatic transformation, however, with the lack of $\gamma_5$ in the Hamiltonian, by removing $h_0 (\bold{k})$ we are also adding an extra symmetry to the system, 
the chiral symmetry.
Thus Kondo insulators involving hybridization with a $\Gamma^J_{1/2}$-doublet
are connected (besides non-adiabatically) to
the fine-tuned Hamiltonian which possess chiral symmetry, e.g.
 \begin{align}
\label{chS}
\gamma_5
{\cal H}_\bold{k}
\gamma_5
&=-{\cal H}_\bold{k}.
\end{align}
The consequences of this connection between $\mathbb{Z}_2$ Kondo insulators 
and  the $\mathbb{Z}$ Hamiltonian (\ref{chS}), in class DIII~\cite{Schnyder2009},
is analysed through the example of tetragonal Kondo insulators 
discussed in this and the following sections.

Hamiltonians in class DIII are characterized by the winding number
  ${N = \int {d^3k\over 48\pi^{2}}
  \epsilon^{ijk}\,
  \text{tr}\left(
  \gamma_5 {\cal H}^{-1}(\partial_{i}{\cal H}){\cal H}^{-1}
  (\partial_{j}{\cal H}){\cal H}^{-1}(\partial_{k}{\cal H})
  \right)}$~\cite{Sato-Ando-2017},
here summation over repeated indices is implicit,
the integral extends over the first Brillouin zone,
$\epsilon^{ijk}$ is the total anti-symmetric Levi-Civita tensor,
 and $\partial_i\equiv \partial_{k_i}$.
The winding number is
related to the Brouwer index of the map
$\bold{k}\mapsto (\bold{h}/|\bold{h}|)(\bold{k})$
 with $\bold{h}^T=(h_1,h_2,h_3,h_4)$, i.e.~\cite{Note-Supplementary-Material}
\begin{align}
\label{bd}
N
 =
 \sum_{\bold k\in\bold h^{-1}(\bold{n}_0)}{\rm sgn} \det
\left(
\partial \bold{h}(\bold{k})
\right),
\end{align} 
where
$\partial \bold{h}$ is the matrix with
elements $(\partial \bold{h})_{ij} =\partial_i h_j$,
and the sum is over
points $\bold{k}$ in the Brillouin zone which map
onto some (arbitrary)
 point $\bold{n}_0$ on the 3-sphere,
 $(\bold h/|\bold h|)(\bold{k})=\bold{n}_0$.
Notice that
the winding counted by \eqref{bd} 
cannot be changed as long as the chiral symmetry is preserved.
That is, the topological properties
are robust against time-reversal symmetry breaking perturbations
that do not violate \eqref{chS}~\footnote{Such perturbations change
 class DIII to AIII without modifying its characterizing topological invariant.}.

\begin{figure}[t]
  \centering
  \includegraphics[width=0.9\linewidth]{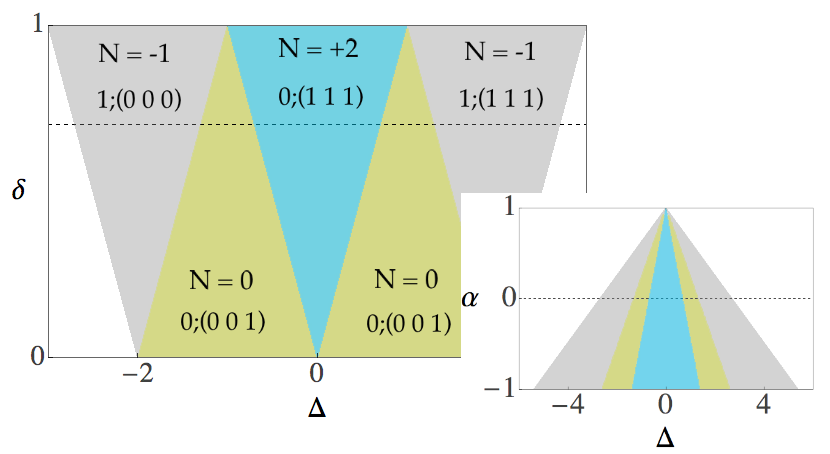}
  \caption{(Color online)
  Phase diagram of $3D$ Kondo insulators with tetragonal symmetry,
  derived from model Eq.~\eqref{HamiltonianDirac} for the $\Gamma^{5/2}_{1/2}$ doublet  
  in the nearest neighbor approximation. Here
   $\delta = t^c_\perp/t_\parallel^c$, $\Delta = (\varepsilon_c - \varepsilon_f)/2(t_\parallel^c-t_\parallel^f)$, and we assumed
  ${v_\perp>0}$, $t_\parallel^f=t_\perp^f=0$.
 Grey regions correspond to strong topological phases with index $\nu_0 = 1$.
 Blue and green regions correspond to 
 weak topological phases with indices $\nu_0;(\nu_1 \nu_2 \nu_3)$ equal to $0;(1 1 1)$ and $0;(0 0 1)$, respectively.
 The winding number $|N|=2/0/1$ in blue/green/gray regions characterises the spin textures of edge states in this system, as discussed in the main text.
White areas are trivial phases with zero in all topological invariants.
 The dashed line indicates the value for $\delta$ used in the inset.
 Inset: Phase-diagram for different (renormalized) band-widths for $f$-electrons
 ${\alpha = t_\parallel^f / t_\parallel^c}$ and fixed $\delta=0.7$.
 The dashed line indicates the value used in the main figure.
}
  \label{fig1Diagram}
\end{figure}

A robust $\Gamma^J_{1/2}$-doublet ground state can only be realize in tetragonal and hexagonal lattices (see section \ref{applications}), and 
the former are the most relevant for application of our results to known Kondo insulators.  
Concentrating then on tetragonal Kondo insulators with a
$\Gamma^{5/2}_{1/2}$-doublet in the ground state,
one finds
(upon using
parameters from the nearest-neighbor
model~\cite{Note-Supplementary-Material})
\begin{align}
\label{windE>0}
\displaystyle
N
&=
\begin{cases}
\, 2\, {\rm sgn}(v_\perp \delta), &
 | \Delta |< |\delta|,
\\
 - {\rm sgn}(v_\perp \delta), \qquad &
 2 - |\delta|  < |\Delta|  <  2 + |\delta|,
\\
 \quad 0, \qquad &
\text{otherwise},
 \end{cases}
\end{align}
where ${\Delta = (\varepsilon_c - \varepsilon_f)/2(t_\parallel^c-t_\parallel^f)}$
is the normalized distance between centers of the bands (i.e. $\varepsilon_c - \varepsilon_f$)
and  $v_\perp$ the hybridization intensity perpendicular to the symmetry plane. 
We also assumed that the anisotropy $\delta$ affects equally the hopping parameters
$t_{\parallel/\perp}^{c,f}$ of $c$- and $f$-electrons    
within/perpendicular to the symmetry plane,
  i.e. $\delta = t^c_\perp / t_\parallel^c$ and $t^f_\perp = \delta t_\parallel^f$.  
The resulting phase-diagram
is shown
in Fig.~\ref{fig1Diagram} for completely localized $f$-electrons, $t_\parallel^f=t_\perp^f=0$.
Just below each winding number signalized in the phase diagram we also show the topological indices $\nu_0 ; (\nu_1 \nu_2 \nu_3)$
following Fu and Kane notation in Ref. [\onlinecite{Fu-Kane-2007}].
The white regions are trivial phases with $0$ in all topological invariants.

\subsection{Improved characterization}
The topological non-trivial regions in Fig. \ref{fig1Diagram} have the surface states ruled by the Fu and Kane indices,
i.e., the indices $\nu_0 = 1$ and $\nu_0 = 0$ set the strong (gray) and weak (blue) topological phases of Kondo insulators respectively, with surface states in all directions, as discussed in [\onlinecite{Dzero2012,Alexandrov2013,Dzero2013}].
The tetragonal structure pushes further away the strong phase from the most relevant parameter regime
and gives rise to an additional topological weak phase $0;(0 0 1)$ (green), with appearing surface states only on those surfaces aligned with $z$ direction.
Notice that increasing the hopping anisotropy ($\delta \rightarrow 0$),
e.g. by application of uniaxial pressure,
induces a phase transition into this new topological weak phase.
Transitions between topological phases
can also occur through correlation induced renormalization of the
$f$-electron dispersion~\footnote{The renormalization factor is e.g. found
from expanding the self-energy
due to interaction between localized $f$-electrons
close to the Fermi surface~\cite{Dzero2012},
or alternatively using slave bosons~\cite{Kusunose2000}
or Gutzwiller's approximation~\cite{Rice1985,Rice1985erratum}}.
The inset shows the phase diagram as a function of
$f$-electron renormalized band-width $t_\parallel^f = \alpha t_\parallel^c$ at fixed $\delta=0.7$
(delimitated by the dashed line in the main figure).

A topologically interesting behavior
is found in the most relevant parameter regime
where the band-width of conduction electrons $t_\parallel^c$ sets
the largest energy scale such that $|\Delta| \ll 1$.
In this phase the winding number provides further description of the surface states,
namely one finds $N \neq 0$ and $0;(1 1 1)$ (i.e., edge states in all surfaces and $|N|=2$).
Projecting the effective tetragonal Kondo Hamiltonians onto the surface states
one finds that the winding number $N$ in Eq. (\ref{bd}) allows us to infer the spin-texture of surface states~\cite{Note-Supplementary-Material}:
specifically it counts the chiralities of the Dirac cones pseudo-spin (spin-texture), i.e.
a weak topological phase
with vanishing winding number indicates an even number of Dirac cones with opposite chiralities,
while finite winding number indicates an even number of Dirac cones with the same chiralities.
Similiar property holds true for the strong topological phases with odd number of Dirac cones,
where the winding number counts the number of Dirac cones left unpaired (pairs of opposite chiralities), 
for example the phase with $|N| = 1$ in the phase diagram of Fig. \ref{fig1Diagram}.

In the context of low neighbor approximation, in Appendix~\ref{cubic} we exemplify with the 4-band model of a cubic structure as described in Ref. [\onlinecite{Vojta2015}].

Previous discussions of the edge states in the general chiral Hamiltonian Eq. (\ref{chS}) are found in Refs.~[\onlinecite{PhysRevB.82.035105,Vayrynen2011,Shen-2013,fradkin_2013}].
In particular, it has been shown the interfaces that break time-reversal symmetry have their gapless edge states replaced by (gapped) non-singular walls and solitons with spin textures protected by the chiral symmetry~\cite{Vayrynen2011}.
Finally, the appearance of spin textures in cubic structures based on the mirror symmetry is discussed in Refs.~[\onlinecite{Vojta2015,PhysRevLett.115.156405}]~\cite{Note-Supplementary-Material}.

\begin{table}[b!]
\begin{tabular}{p{2.3cm}|p{3cm}|p{2cm}}
\hline
{\bf Crystal-Field} & $\,\,\,\,$ {\bf Point-Group} & $\,\,${\bf Multiplet}
\\
\hline
\hline
$\,\,$ tetragonal
&
$\,\,$ C$_4$, S$_4$, C$_{4h}$, D$_4$,$\,\,\,\,\,$
$\,\,$\hspace{.5cm} C$_{4v}$, D$_{2d}$, D$_{4h}$
&
$\,$ 5/2
\tabularnewline
\hline
$\,\,$ hexagonal
&
$\,\,$ C$_6$, C$_{3h}$, C$_{6h}$, D$_{3h}$
&
$\,$ 5/2, 7/2
\tabularnewline
\hline
\end{tabular}
\caption{Point group symmetries which separate a pure
$\Gamma^J_{1/2}$ Kramers'-doublet with lowest projection of
angular momentum from the spin-orbit multiplets $J=5/2$ and $7/2$.
This Kramers' doublet can split from the $5/2$-sextet
in all of the seven point group symmetries of the tetragonal
lattice, or four out of the seven point group symmetries of the hexagonal
lattice. In case of the
$7/2$-octet the Kramers' doublet with lowest projection of
angular momentum can only split in four out of point group symmetries
of the hexagonal lattice.
}
\label{table1}
\end{table}

\section{Applications}
\label{applications}
Candidate-compounds for topological Kondo insulators
are formed from
magnetic ions with ground states involving odd-parity orbitals.
Concentrating onto the rare-earth 3$^+$-ions with partially filled $4f$ shell,
 Ce-, Sm-, and Yb-based materials are of potential interest.
 The ground state Kramers' doublet in case of the former two compounds arises 
 from the 5/2-sextet and in case of the latter from the 7/2-octet. 
 The necessary requirement for the appearance of a Kondo insulator as discussed here
 is then a
crystal field which stabilizes the $\Gamma^J_{1/2}$-doublet
in the ground state.
Looking at representations of all possible point groups and their basis 
functions~\cite{Bradley-Cracknell-1972}, we notice that from the $f$-electron multiplets 
a pure $\Gamma^J_{1/2}$-doublet 
only separates in tetragonal or hexagonal crystal symmetries.
Specifically, the $\Gamma^{5/2}_{1/2}$-doublet is allowed
as one possible ground state in all tetragonal lattices and some of the hexagonal lattices, while
the $\Gamma^{7/2}_{1/2}$-doublet can only be a ground state in some of the hexagonal lattices.
Table~\ref{table1} summarizes the possible point group symmetries
 of lattices allowing for a Kondo insulator derived from the $5/2$- and
 $7/2$-multiplets, respectively.

We conclude that Ce- and Sm-based Kondo insulators can only have $\Gamma^J_{1/2}$-doublet ground state
in tetragonal or hexagonal structures. 
One specific Ce-compound with tetragonal point group symmetry D$_{2d}$, 
to which our above analysis applies is 
CeRu$_4$Sn$_6$~\cite{Paschen2010}. 
Recent x-ray spectroscopy experiments in combination with
 band structure calculations indicate that the 
$\Gamma^{5/2}_{1/2}$-doublet is the lowest energy state
 and inversion of bands occurs~\cite{Sundermann,Sundermann2017,Note-DMFT-Ce-compound}.
Moreover, the $4f$ occupancy near to integer value $n_f \sim 1$ 
and the low dispersive $f$-band, 
put this material into the topologically interesting region $|N|=2$
of the phase diagram, Fig.~\ref{fig1Diagram}.
All known Sm-based Kondo insulators, on the other hand, have cubic symmetry and 
our analysis does not apply. 
Finally, Kondo insulators based on Yb can only 
exhibit $\Gamma^J_{1/2}$-doublet ground state in the four hexagonal
symmetries indicated in table~\ref{table1}.
Among the established Yb based Kondo insulators there is none with hexagonal symmetry, i.e. 
these compounds can at most be realized in low-order neighbor approximation.
Recently, an interesting Yb compound with hexagonal symmetry, YbNi$_3$$X_9$ ($X =$ Al, Ga),
has been synthesized, but it appears to be metallic~\cite{Yamashita-2012}.

Besides the rare-earth elements, Kondo insulators may also be find in metal transition elements,
for instance the new Iridium-based compound Sr$_2$IrO$_4$\cite{PhysRevLett.101.076402} has a narrow $5d$ band from Ir which hybridizes with $4p$ band from Oxygen.
It also shows a $\Gamma^J_{1/2}$-doublet ground state and has a tetragonal lattice structure.

\section{Discussion}
\label{discussion}
We have studied $3D$ Kondo insulators,
where a wide conduction band
hybridizes with a degenerate Kramers' doublet of localized $f$-electrons.
We have shown that in cases where the doublet is that of lowest angular momentum
projection, $m_J=\pm1/2$,
the symmetry of orbitals involved allows for an improved
characterization of the topological properties.
The existence of an underlying low-energy effective field theory for three-dimensional ($3D$) topological insulators
guarantees that the  electromagnetic and thermal responses are associated with topological invariants.
Here the clean system is connected with a fine-tuned Hamiltonian in class DIII, which in turn is characterized by the $\mathbb Z$-invariant.
In this case, the winding number distinguishes the chirality of Dirac cones at the surfaces, 
providing further informations about the edge states.
Shiozaki et al.~\cite{ShiozakiPRL2013} have investigated the possible realizations of physical quantities
which distinguish the $\mathbb{Z}_{2}$ properties from the $\mathbb{Z}$ ones.

As an example, in cases where the band width of conduction electrons sets the largest energy scale,
the tetragonal topological Kondo insulator is in a nontrivial
phase with winding number $|N|=2$, which means that we have two Dirac cones with the same chirality at each surface.
Moreover, the phase diagram of this Kondo insulator shows a new weak topological phase
when increasing the hopping anisotropy from the cubic to tetragonal structure.
 When other than the $m_J=\pm1/2$-doublet is involved in the hybridization
 or when the crystalline field is other than one of those listed in table~\ref{table1},
such system may only appear
in a low order neighbor approximation.

Relevant crystal structures for this work are thus those which allow
for a pure $\Gamma^J_{1/2}$ doublet in the ground state.
This implies that topological Kondo insulators involving Kramers' doublets from
the $7/2$ spin-orbit octet can only exist in some of the crystalline hexagonal lattices (see table~\ref{table1}).
Kondo insulators forming from
hybridization with a  Kramers' doublets from
the $5/2$-sextet, on the other hand, can exist in all tetragonal
and some of the hexagonal lattices.
In practice, the crystal field splitting may
not be strong enough to separate the ground state and
(anisotropic) pressure
may help to stabilize a topological phase.
Finally, we have discussed several implications for the rare-earth compounds.

The authors thank the extensive discussions with T.Micklitz
and T.O.P. thanks helpful comments from P.Aynajian and S.Kirchner.
This work was supported by Brazilian Research Agencies CNPq and FAPERJ,
and Chinese Agency NSFC under grant numbers 11750110429 and U1530401,
and Chinese Research Center CSRC.

\bibliographystyle{apsrev4-1}
\bibliography{refs}

\begin{thebibliography}{48}
\bibitem [{Note4({\natexlab{a}})}]{Note-Supplementary-Material}%
  \BibitemOpen
  \bibinfo {note} {See appendix, where we review the Brouwer index, give
  details on hybridization-matrices, discuss the tetragonal Kondo insulator,
  and exemplify a non-zero winding number with spin textures in cubic
  structures.}\BibitemShut {Stop}%
\end{thebibliography}


\appendix

\section{Winding number and Brouwer index}
\label{App1}

For convenience of the reader we here review the calculation of the winding number,
 $N 
 = \int {d^3k\over 48\pi^{2}}
  \epsilon^{ijk}\,
  \text{tr}\left(
  \gamma_5 {\cal H}^{-1}(\partial_{i}{\cal H}){\cal H}^{-1}
  (\partial_{j}{\cal H}){\cal H}^{-1}(\partial_{k}{\cal H})
  \right)$,
via the Brouwer degree, Eq.~(7) in the main text. 
 Starting out from the Hamiltonian matrix
${\cal H}(\bold{k})= \sum_{i=1}^4 h_i(\bold{k})\gamma_i$ and the chiral symmetry 
operator
$\gamma_5=\gamma_1\gamma_2\gamma_3\gamma_4$,
we use anti-commutation relation, $\{\gamma_a,\gamma_b\}=2\delta_{ab}$,
to simplify terms, e.g.
$h_ah_b\gamma_a\gamma_b=\tfrac{1}{2}h_ah_b(\gamma_a\gamma_b+\gamma_b\gamma_a)
=h_ah_b\delta_{ab}\equiv|h|^2$, etc. 
One then arrives at 
$N
= 
\int {d^3k\over 48\pi^2} 
f^{ijk}_{abcd}
{h_a\over |h|^4}
 (\partial_ih_b)(\partial_jh_c)(\partial_kh_d)$, 
with $f^{ijk}_{abcd}=\epsilon^{ijk}{\rm tr}
\left(\gamma_5
 \gamma_a \gamma_b \gamma_c \gamma_d
\right)$, which can be cast into the form
\begin{align}
\label{pullback}
N=
\frac{1}{12\pi^2} \int \,
{1\over |h|^4}\epsilon^{a b c d  }
h_a
dh_b \wedge d h_c \wedge dh_d.
\end{align}
Eq.~\eqref{pullback} is the pull-back 
of the (normalized) volume form on the three-sphere,
i.e. $N=\int h^*\omega_{S^3}$, where
$\omega_{S^3}
=
{1\over {\rm vol}(S^3)}{1\over |k|^4}  i_{k^a \bold{e}_a}(dk_1\wedge dk_2\wedge dk_3\wedge dk_4)$.
The latter can be calculated from the 
Brouwer degree of the map
$\bold{h}/|\bold{h}|: T^3\to S^3$, $\bold{k}\mapsto (\bold{h}/|\bold{h}|)(\bold{k})$
 with $T^3$ the $3D$ Brillouin zone torus and
 $\bold{h}^T=(h_1,h_2,h_3,h_4)$. The Brouwer degree
 counts the number of intersections
of a ray through the origin and the oriented surface spanned by the map,
 as discussed in the main text.

\section{Hybridization matrix for the $5/2$-doublets in tetragonal crystal field}
\label{apphybridization}

\begin{table}[t!]
\begin{tabular}{c|c|c|c|c}
\hline
$V_{n-\frac{1}{2}}$
& $h_{1}^{n}$  
& $h_{2}^{n}$ 
& $h_{3}^{n}$ 
& $h_{5}^{n}$  
\tabularnewline
\hline
\hline
$n=1$  & $v_\parallel\sin\left(k_{x}\right)$ & $v_\parallel\sin\left(k_{y}\right)$ & $-2 v_\perp \sin\left(k_{z}\right)$ & $0$
\tabularnewline
\hline
$n=2$  & $0$ & $0$ & $v_\parallel \sin\left(k_{y}\right)$ & $v_\parallel \sin\left(k_{x}\right)$ \tabularnewline
\hline
$n=3$ & $-v_\parallel \sin\left(k_{x}\right)$ & $v_\parallel \sin\left(k_{y}\right)$ & $0$ & $0$ \tabularnewline
\hline
\end{tabular}
\caption{
Coefficient functions $h_{1,2,3,5}^{n}\left(\bold{k}\right)$ 
parametrizing the hybridization matrix $V_{n-\frac{1}{2}}\left(\bold{k}\right)$
for the $m_J=1/2,3/2$, and $5/2$ doublets of the $J=5/2$-sextet in the 
nearest-neighbor approximation of a tetragonal lattice.
Here  $v_\parallel \equiv v(r^1_{x,y})$ and
${v_\perp \equiv v(r^1_{z})}$
are the hybridization intensities within and perpendicular to the symmetry ($x,y$)-plane,
respectively.}
\label{hybridization first nn}
\end{table}

Let us recall that matrices Eqs.~(3)  and (5) in the main text describe the hybridization
\begin{align}
\hat H_V(\bold{k})
&=
\begin{pmatrix}c_{\bold{k}\uparrow}^{\dagger} 
& c_{\bold{k}\downarrow}^{\dagger}\end{pmatrix}
V_{m_J}(\bold{k})
\begin{pmatrix}f_{\bold{k},+m_J}\\
f_{\bold{k},-m_J}
\end{pmatrix},
\end{align}
where $m_J = n-\frac{1}{2}$ from Eq. (3) in the main text.
Concentrating then on Kramers' doublets separating from the $J=5/2$-sextet  and a tetragonal symmetry, 
we find in the nearest neighbor approximation 
the coefficient functions $h_{1,2,3,5}^{n}\left(\bold{k}\right)$ summarized in table~\ref{hybridization first nn} 
(see Eq. (4) in the main text). 
This situation was also considered in Ref. [\onlinecite{PhysRevB.85.125128}] up to first next neighbour approximation.

We notice that 
in the nearest neighbor approximation hybridization with $m_J=3/2$ and $5/2$ doublets 
does not open a gap in the spectrum and 
the system remains metallic. 
As discussed in the main text, vanishing coefficient functions for $m_J=5/2$ and $7/2$ doublets 
are related to the specific values of spherical harmonics $Y_3^{\pm 2}$ at the angles of the nearest neighbor directions, here 
in the tetragonal lattice. In case of the $m_J=1/2$ doublet, on the other hand, vanishing of $h_5$ is a consequence of the 
rotational symmetry of $Y_3^0$, and independent of the nearest neighbor approximation.

Accounting for next-nearest neighbor contributions, a (small) gap also opens 
in case of hybridization with $m_J=3/2$ and $5/2$ doublets,
 as can be seen from
table~\ref{hybridization second nn}, where we summarize coefficient functions now including 
next-nearest neighbor contributions.
Here we defined
$F_{a} = F_{( y - z)}$,
$F_{a}' = F_{-( y - z)}$,
$F_{b} = F_{( x - z)}$,
and
$F_{b}' = F_{-( x - z)}$,
with
$F_{\pm ( i \pm j)} = v_\parallel + v_{2\parallel} (1 \pm 2 C_{i \pm j} -\sqrt{2} \cos (k_z) )$, 
$C_{i \pm j} = \sqrt{2} \left ( \cos(k_i) \pm \cos (k_j) \right )$
($i,j = x,y,z$), and $F_{c} = 2 v_\perp + v_{2\perp} ( 2 - C_{x+y})$,
$F_c' = v_{2\perp} C_{x-y}$.
Here $v_{\parallel/\perp}$ and $v_{2\parallel/2\perp}$ 
are the hybridization intensities within/perpendicular to the symmetry plane 
for first- and second-nearest neighbor sites, respectively
(as also used in Eq.~(4) of the main text).
Notice that, in this order of hopping approximation, effective models for $m_J=3/2$ and $5/2$-doublets 
also show a chiral symmetry, i.e. $\gamma_1$ and $\gamma_5$, respectively. 
Including, however, contributions from third-nearest neighbors all coefficient functions become 
non-vanishing 
in case of $m_J=3/2$ and $5/2$-doublets. Only in case of the $m_J=1/2$ doublet $h_5$ 
remains zero.
Finally, a discussion similar to the above applies to coefficient functions parametrizing 
hybridization with doublets from the $J=7/2$-octet.

\begin{table}[t]
\begin{tabular}{c|c|c|c|c}
\hline
$V_{n-\frac{1}{2}}$  
& $h_{1}^{n}$ & $h_{2}^{n}$ & $h_{3}^{n}$
& $h_{5}^{n}$
\tabularnewline
\hline
\hline
$n=1$  & $F_{a}\sin\left(k_{x}\right)$ & $F_{b}\sin\left(k_{y}\right)$ & $-F_{c}\sin\left(k_{z}\right)$& $0$ \tabularnewline
\hline
$n=2$  & $0$ & $5F_{c}'\sin\left(k_{z}\right)$ & $F_{b}\sin\left(k_{y}\right)$& $F_{a}\sin\left(k_{x}\right)$ \tabularnewline
\hline
$n=3$  & $-F_{a}'\sin\left(k_{x}\right)$ & $F_{b}'\sin\left(k_{y}\right)$ & $-F_{c}'\sin\left(k_{z}\right)$& $0$ \tabularnewline
\hline
\end{tabular}
\caption{
Coefficient functions $h_{1,2,3,5}^{n}\left(\bold{k}\right)$ 
as in table~\ref{hybridization first nn}, now including
next-nearest neighbor contributions.}
\label{hybridization second nn}
\end{table}

\section{Tetragonal Kondo insulator}
\label{tetraKI}

We here focus on a tetragonal Kondo insulator  
with the $\Gamma^{5/2}_{1/2}$-doublet in the ground state and 
calculate the winding number from the nearest neighbor model.
Dispersion relations for conduction and (nearly) localized electrons then read 
$\varepsilon_{\bold{k}}^{c,f} = \varepsilon_{c,f} + 2t_\parallel^{c,f} (\cos (k_x) + \cos (k_y)) + 2t^{c,f}_\perp \cos (k_z)$,
where $\varepsilon_{c,f}$ are the corresponding band-centers 
and 
$t_{\parallel/\perp}^{c,f}$ hopping parameters within/perpendicular to 
the symmetry plane of the tetragonal structure. 
Dispersion relations define coefficients $h_{0,4}(\bold{k})$ (see Eq. (2) in the main text) and 
coefficients $h_{1,2,3,5}^{n}\left(\bold{k}\right)$
are taken e.g. from table~\ref{hybridization first nn} or \ref{hybridization second nn} of the previous section. 
With these functions Eq.~(7) in the main text reads
\begin{align}
\label{Winding Gamma1}
N 
&= 
  \sum_{\bold k\in\bold h^{-1}(\bold{n}_0)} {\rm sgn} (-F_a F_b F_c \cos k_x  \cos k_y \cos k_z),
\end{align}
where $F_{a,b,c}$ have been discussed in the previous section. 
To evaluate the sum~\eqref{Winding Gamma1}, it is then convenient to choose $\bold{n}_0 = h_4(0)\bold{e}_4$ 
whose pre-image, $\bold{h}^{-1}(\bold{n}_0)$, are the eight time-reversal invariant points in the Brillouin zone. 
The result of this calculation is given in Eq.~(8) of the main text.

\section{Surface-states spin texture in the tetragonal Kondo insulator}
\label{texturesTetraKI}

The translational invariant tetragonal Kondo insulators allows for
a characterization in terms of winding number, as described in Eq.
(7) in the main text and appendix A. Here we apply the projection
method to derive the surface Hamiltonian. Let us consider those time-reversal
invariant momenta points $\boldsymbol{k}_{0}$, as described in the
previous section. In their vicinity the Hamiltonian reads ${\cal H}\left(\boldsymbol{k}\right)=\sum_{i=1}^{3}v_{i}k_{i}\sigma_{i}\tau_{x}+m\tau_{z},$
with parameters $v_{i}$ and $m$ functions of $\boldsymbol{k}_{0}$.
As an illustration we consider the surface Hamiltonian at $z=\pm L/2$,
where $L$ is the $z$-direction system size. Since translational
invariance is broken in $z$-direction we substitute $k_{z}\rightarrow-i\partial_{z}$
and the zero energies eigenfunctions are obtaining by
\[
\left(\tau_{z}\left[m-\left(P_{+}-P_{-}\right)v_{z}\partial_{z}\right]+\sum_{i=1}^{2}v_{i}k_{i}\sigma_{i}\tau_{x}\right)\psi_{\boldsymbol{k}}\left(z\right)=0,
\]
where we have introduced the projection operators $P_{\pm}=\frac{1}{2}\left(\mathbb{I}\pm\sigma_{z}\tau_{y}\right)$.
The spatially dependent part of the Schr\"odinger
equation, with $\psi_{\boldsymbol{k}}\left(z\right)=\psi\left(z\right)\psi\left(\boldsymbol{k}\right)$,
is solved by eigenfunctions of $P_{\pm}$, that is, introducing $P_{\pm}\psi^{\pm}=\pm\psi^{\pm}$
the $z$-coordinate dependent part reads
\[
\psi\left(z\right)=e^{\frac{m}{v_{z}}z}\psi^{+}+e^{-\frac{m}{v_{z}}z}\psi^{-}.
\]
Depending on the sign of $\text{sgn}\left(m/v_{z}\right)=\pm$ the
first/second contribution accounts for the wave-functions exponentially
localized at $z=\mp L/2$. Concentrating on either one of the surfaces
we project the $\boldsymbol{k}$-dependent part on the corresponding
eigenspace $H^{\pm}\equiv P_{\pm}H\left(\boldsymbol{k}\right)P_{\pm}$.
In order to find an explicit expression it is convenient to introduce
$U\equiv e^{i\frac{\pi}{4}\tau_{x}}$ such that the surface Hamiltonians
are written in the rotated basis $H_{U}^{\pm}=U^{\dagger}H^{\pm}U$,
explicitly
\begin{align*}
H_{U}^{+} & =\begin{pmatrix}0 & 0 & 0 & v_{x}k_{x}-iv_{y}k_{y}\\
0 & 0 & 0 & 0\\
0 & 0 & 0 & 0\\
v_{x}k_{x}+iv_{y}k_{y} & 0 & 0 & 0
\end{pmatrix},\\
H_{U}^{-} & =\begin{pmatrix}0 & 0 & 0 & 0\\
0 & 0 & v_{x}k_{x}+iv_{y}k_{y} & 0\\
0 & v_{x}k_{x}-iv_{y}k_{y} & 0 & 0\\
0 & 0 & 0 & 0
\end{pmatrix}.
\end{align*}
From this result we noticed that each of the two Hamiltonian describes
a given surface depending on $\text{sgn}\left(m\left(\boldsymbol{k}_{0}\right)v_{z}\left(\boldsymbol{k}_{0}\right)\right)$.
The surface Hamiltonians on opposite surfaces have opposite chiralities,
i.e. $\text{ch}_{+}=\text{sgn}\left(v_{x}v_{y}\right)$ and $\text{ch}_{+}=-\text{sgn}\left(v_{x}v_{y}\right)$.
Thus, the chirality of the surface states at a given surface is fixed
by the product $\text{ch}=\text{sgn}\left(mv_{x}v_{y}v_{z}\right)$.
Coming back to our example in the previous section, the sum is over
time-reversal invariant momenta where band inversion occurs, i.e.
$m\left(\boldsymbol{k}_{0}\right)<0$. Having fixed the Brouwer's
formula $\boldsymbol{n}_{0}=h_{4}\left(0\right)\bold{e}_{4}$ we noticed
that each summand is related to the chirality of surface states such
that
\[
N=\sum_{\boldsymbol{k}_{0}}\text{sgn}\left(-v_{x}\left(\boldsymbol{k}_{0}\right)v_{y}\left(\boldsymbol{k}_{0}\right)v_{z}\left(\boldsymbol{k}_{0}\right)\right).
\]
The absolute value of the winding accounts for the total chirality
of surface states when present on a given surface. The latter is a
well defined quantity, i.e. independent of the surface one looks at.

\section{cubic Kondo insulator at low neighbor hopping approximation}
\label{cubic}

As an example of low neighbor hopping approximation,
we apply our calculations to the $4$-band model of a cubic structure
as described in Ref. [\onlinecite{Vojta2015}]. According to our notation, their Hamiltonian
can be rewritten as 
$\varepsilon_{\boldsymbol{k}}^{c,f}=\epsilon_{0}^{c,f}-2t_{c,f}\eta_{1}^{c,f}\left(c_{x}+c_{y}+c_{z}\right)-4t_{c,f}\eta_{2}^{c,f}\left(c_{x}c_{y}+c_{y}c_{z}+c_{z}c_{x}\right)$,
where $\varepsilon_{\boldsymbol{k}}^{c,f}$ are dispersion relations
for the conducting and localized bands, $\epsilon_{0}^{c,f}$ are
the corresponding band-centers, and $t_{c,f}\eta_{1}^{c,f}$ and $t_{c,f}\eta_{2}^{c,f}$
are the band-width for first and second nearest neighbors, respectively,
finally $c_{i}=\cos\left(k_{i}\right)$ with $i=x,y,z$. Using notation
of Table \ref{hybridization second nn} with $n=1$, the hybridization elements have their coefficient
functions as $F_{a}=-2V\left(\eta^{v1}+\eta^{v2}\left(c_{y}+c_{z}\right)\right)$,
$F_{b}=-2V\left(\eta^{v1}+\eta^{v2}\left(c_{x}+c_{z}\right)\right)$
and $F_{c}=2V\left(\eta^{v1}+\eta^{v2}\left(c_{y}+c_{x}\right)\right)$,
where $V\eta^{v1}$ and $V\eta^{v2}$ are the hybridization amplitudes
for nearest and next-nearest neighbor hoppings, respectively. Finally,
the parameters were set to $\epsilon_{0}^{f}-\epsilon_{0}^{c}=-2eV$,
$t_{c}=1eV$, $t_{f}=0.003eV$, $\eta_{1}^{c}=\eta_{1}^{f}=1$, $\eta_{2}^{c}=\eta_{2}^{f}=-0.5$,
$V\eta^{v1}=0.2eV$, and $V\eta^{v2}=0$. 

Now we intend to calculate the winding number according to Eq. \ref{Winding Gamma1},
where we evaluate the sum by choosing $\boldsymbol{n}_{0}=-h_{4}\left(0\right)\bold{e}_{4}$,
whose pre-image are the eight time-reversal invariant points in the
Brillouin zone. The result of this calculation is $N=+3$, which characterizes
the three Dirac cones with the same pseudo-spin chirality in Fig.
3(a) in Ref. [\onlinecite{Vojta2015}].

\end{document}